\documentclass[%
 aip,
 amsmath,amssymb,
 reprint,%
]{revtex4-1}

\usepackage{graphicx}
\usepackage{dcolumn}
\usepackage{bm}

\usepackage[utf8]{inputenc}
\usepackage[T1]{fontenc}
\usepackage{mathptmx}
\usepackage{etoolbox}

\usepackage{lineno}
\usepackage{comment}
\usepackage{xcolor}
\usepackage{acronym} 
\usepackage{pdfcomment}
\usepackage{hyperref}
\usepackage{braket}
\usepackage{subcaption}
\usepackage{caption}
\usepackage{siunitx}
\usepackage{adjustbox}
\usepackage{makecell}
\usepackage{multirow}
\usepackage{wrapfig}
\usepackage{array}
\usepackage{ragged2e}
\usepackage{lipsum}
\usepackage{epstopdf}
\usepackage[english]{babel}
\usepackage{enumitem}

\newcommand{\ed}[1]{{\color{red} #1}}

\renewcommand{\arraystretch}{1.6} 

\DeclareCaptionJustification{justified}{\justifying}

\makeatletter
\def\@email#1#2{%
 \endgroup
 \patchcmd{\titleblock@produce}
  {\frontmatter@RRAPformat}
  {\frontmatter@RRAPformat{\produce@RRAP{*#1\href{mailto:#2}{#2}}}\frontmatter@RRAPformat}
  {}{}
}%

\begin{document}

\pagenumbering{arabic}

\preprint{AIP/123-QED}

\title{Quantum key distribution over a metropolitan network using an integrated photonics based prototype}


\author{Maria Ana Pereira}
 \affiliation{Quantum Technologies Group, Department of Applied Physics, Université de Genève, Switzerland}
 \email{maria.dematosafonsopereira@unige.ch}
 
\author{Giulio Gualandi}%
\affiliation{Department of Physics, Politecnico di Milano, Italy}
\affiliation{ Institute for Photonics and Nanotechnologies, CNR, Italy}

\author{Rebecka Sax}
\affiliation{Quantum Technologies Group, Department of Applied Physics, Université de Genève, Switzerland}

\author{Alberto Boaron}
\affiliation{Quantum Technologies Group, Department of Applied Physics, Université de Genève, Switzerland}

\author{Raphaël Houlmann}
\affiliation{Quantum Technologies Group, Department of Applied Physics, Université de Genève, Switzerland}

\author{Roberto Osellame}
\affiliation{ Institute for Photonics and Nanotechnologies, CNR, Italy}
\affiliation{Department of Physics, Politecnico di Milano, Italy}

\author{Rob Thew}
\affiliation{Quantum Technologies Group, Department of Applied Physics, Université de Genève, Switzerland}

\author{Hugo Zbinden}
\affiliation{Quantum Technologies Group, Department of Applied Physics, Université de Genève, Switzerland}

\date{\today}

\begin{abstract}
Transitioning \ac{QKD} towards industrial-scale deployment requires the development of practical, stable, resilient and cost-effective hardware that can be manufactured at large scales.
In this work we present a high-speed (\qty{1.25}{\giga\hertz}), field-deployable \ac{QKD} prototype based on integrated photonics, consolidated into standard 19-inch rack compatible units.
This prototype leverages integrated photonics to address the requirements for autonomous long-term stability in metropolitan settings. The architecture demonstrates the potential for simplifying metropolitan links by removing the need for chromatic dispersion compensation up to \qty{100}{\kilo\meter} through optimized laser and detection parameters.
We demonstrate continuous key exchange over a deployed metropolitan optical fiber link, where the prototype maintained stable, uninterrupted operation across a measurement spanning more than 12 day-night cycles without manual intervention. Furthermore, the system's performance limits were characterized by extending the channel with fiber spools and extra attenuation on the receiver side, demonstrating continuous key exchange up to \qty{105}{\kilo\meter} without chromatic dispersion compensation.

\end{abstract}
\maketitle

\acrodef{QKD}{Quantum Key Distribution}
\acrodef{PIC}{Photonic Integrated Circuit}
\acrodef{QC}{Quantum Channel}
\acrodef{MZI}{Mach–Zehnder interferometer}
\acrodef{IM}{Intensity Modulator}
\acrodef{DCF}{Dispersion Compensating Fiber}
\acrodef{QC}{Quantum Channel}
\acrodef{RF}{Radio Frequency}
\acrodef{VOA}{Variable Optical Attenuator}
\acrodef{FPGA}{Field Programmable Gate Array}
\acrodef{QBER}{Quantum Bit Error Rate}
\acrodef{TEC}{Thermo Electric Cooler}
\acrodef{EIC}{Electronic Integrated Circuit}
\acrodef{FWHM}{Full Width Half Maximum}
\acrodef{MMI}{Multimode Interference}
\acrodef{BS}{Beam Splitter}

\acrodef{EOPS}{Electro-Optical Phase-Shifter}
\acrodef{TOPS}{Thermo-Optical Phase-Shifter}
\acrodef{NFAD}{Negative Feedback Avalanche Diode}
\acrodef{PDE}{Photon Detection Efficiency}
\acrodef{SKR}{Secret Key Rate}
\acrodef{SMF}{Single Mode Fiber}
\acrodef{DCR}{Dark Count Rate}
\acrodef{SPAD}{Single Photon Avalanche Diode}
\acrodef{APP}{Afterpulsing Probability}
\acrodef{SOI}{Silicon On Oxide}
\acrodef{BiCMOS}{Bipolar CMOS}
\acrodef{DAC}{Digital-to-Analog-Converter}
\acrodef{PCB}{Printed Circuit Board}
\acrodef{AA}{Absorption Attenuator}
\acrodef{SNSPD}{Superconducting Nanowire Single Photon Detector}
\acrodef{TDC}{Time-to-Digital Converter}
\acrodef{PID}{Proportional-Integral-Derivative}
\acrodef{SMF}{Single Mode Fiber}
\acrodef{SOP}{State of Polarization}

    
Although \ac{QKD} has been established as a standard for information-theoretic security for more than two decades\cite{Pirandola2020-mb}, its widespread adoption across networks has been constrained by hardware limitations. 
Previous works in field-deployability have already demonstrated that long-term autonomous stability is achievable over existing communication infrastructures \cite{Chen2025-cw}. However, early implementations often relied on bulky, expensive components that are difficult to scale within dense telecommunication environments.
Consequently, for over a decade, the focus has shifted from fundamental protocol demonstrations to the realization of scalable architectures \cite{RevModPhys.92.025002}. Among the various approaches to address these scalability challenges, \acp{PIC} have emerged as a promising solution by allowing the consolidation of multiple optical components on a single sub-centimeter chip, eliminating the need for discrete bulk optical components that are sensitive to mechanical perturbations and thermal gradients. The resulting compact form factor and low-power electrical consumption in the order of \unit{\milli\watt}, compared to hundreds of \unit{\milli\watt} in bulk systems, allow the implementation of more practical and environmentally robust modules suitable for real-world telecommunication infrastructures \cite{Labonte2024-bp}.

In this paper, we present the design, implementation and characterization of a \ac{PIC} based, field-deployable \ac{QKD} system based on time-bin encoding.
Our system was designed with scalability, manufacturing efficiency and cost in mind. Using standardized footprints for the receiver and transmitter, it can accommodate a wide range of link distances and losses. 
The system has successfully performed key exchanges on more than \qty{4}{\kilo\meter} of metropolitan optical fiber with stable operation tested for around 12 day-night cycles.
    
\begin{figure}[t!]
    \centering
    \captionsetup{justification=justified, singlelinecheck=false, format=plain}
    \includegraphics[width=1\linewidth]{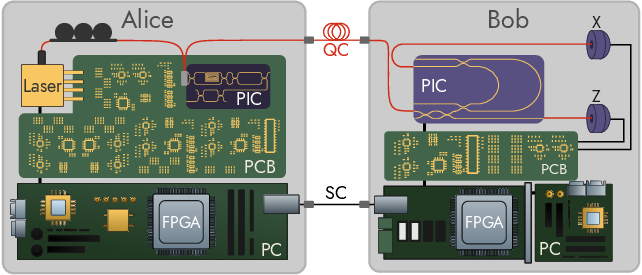}
    \caption{Schematic of the QKD system between a transmitter, Alice, and a receiver, Bob. Alice's setup includes a laser source, a polarization controller and a photonic integrated chip (PIC). The quantum channel (QC) transmits the quantum states to Bob that consists of a PIC and single-photon detectors (Z and X-basis measurements). Both parties are supported by printed circuit boards (PCB), on-board computers (PC) and Field-Programmable Gate Arrays (FPGA) for control and processing. A service channel (SC) is used for clock synchronization, basis reconciliation, error correction, and privacy amplification}
    \label{fig:QKD_Prototype_Setup}
\end{figure}

The protocol used in this work is the 
3-state time-bin encoded BB84 with one decoy state\cite{Boaron_2018}. The full security analysis of the protocol used is detailed in the works of Rusca\cite{Rusca2018-ch,Rusca_SecurityProof}.
The prototype implementing this protocol is shown in Fig.\ref{fig:QKD_Prototype_Setup}. 

The prototype is divided into two main parts: Alice's transmitter unit, consisting of a gain-switched laser, a polarization controller, the transmitter integrated chip, a custom electronic board, a computer and a \ac{FPGA} chip; and Bob's receiver, comprising the receiver integrated chip, two avalanche single-photon detectors, custom electronic board, a computer and an \ac{FPGA}. The transmitter fits within a 1U 19-inch rack box while the receiver fits within a 3U 19-inch rack box. The additional 2U height is required to accommodate the 90-degree connection between the \ac{FPGA} (\textit{AMD Kintex KC705 Evaluation Kit}) and the on-board computer. 

Building on the \ac{PIC}-based \ac{QKD} experiment of Sax\cite{Sax:23}, we introduced design improvements that transformed the integrated photonics systems from a proof-of-principle demonstration into a deployable prototype, with both the transmitter and receiver units redesigned with an emphasis on integration and miniaturization. The transmitter merged clock generation, RF-signal generation, laser driving circuitry, power supplies, and temperature control into a single \ac{PCB}. 
At the receiver, detector units were significantly scaled down, shifting from a Stirling-cooler implementation (\qty{2.3e-2}{\meter\cubed}) to a peltier-cooled approach (\qty{4.6e-4}{\meter\cubed}).

We also reduced the system's operating frequency to \qty{1.25}{\giga\hertz} (from \qty{2.5}{\giga\hertz} \cite{Sax:23}) to mitigate the effects of chromatic dispersion. At \qty{1.25}{\giga\hertz}  each time-bin is \qty{400}{\pico\second} rather than \qty{200}{\pico\second}, making the measurement less susceptible to dispersion-induced temporal broadening (for the same laser temporal profile) and reducing the system's sensitivity to detector jitter.
By making the system more dispersion-tolerant, we eliminate the need to design and adjust the length of \ac{DCF} for any given distance. So, while halving the repetition rate reduces the achievable raw key rate, the resulting improvement in \ac{QBER} and in dispersion tolerance can result in a more effective secret key extraction, especially in a regime where detector saturation is already expected.

A custom \ac{PCB} was designed in house with all the necessary electronics for the optical control of the prototype. 
An \ac{FPGA} interface stage converts the differential signals from the \ac{FPGA} to single-ended SMA outputs to be used as the laser clock and an auxiliary clock. The laser clock features a programmable delay of up to \qty{100}{\pico\second} in steps of \qty{3}{\pico\second} that allows the user to precisely align temporally the optical pulse train and the electronic modulation signal.

The laser used is a high-bandwidth distributed feedback laser (Gooch \& Housego's AA0701 series - \qty{1550.12}{\nano\meter} at \qty{25}{\degreeCelsius}), with a central wavelength that can be thermally tuned to match the requirements of the setup. It is mounted directly on the control \ac{PCB} that provides both the DC bias current and the processed \ac{RF} \ac{FPGA} clock signal (converted to single-end and amplified) that is used as a modulation signal to generate the gain-switched optical pulses. 

The \ac{PIC} and \ac{EIC} are maintained at \qty{40}{\degreeCelsius} using a Peltier cooling stage controlled by a \ac{TEC} controller (\textit{Thorlabs MTD415T}). A microcontroller serves as the central control hub, interfacing with the \ac{EIC} to provide the control signals for parameter adjustments in the \ac{PIC}.

The integrated transmitter comprises a \ac{PIC} and an auxiliary \ac{EIC}, both fabricated using standard silicon photonic technology and designed by Sicoya GmbH. Details of the fabrication are provided in Sax\cite{Sax:23}. The fabricated chips had footprints of \qty{4.50}{\milli\meter} $\times$ \qty{1.10}{\milli\meter} and \qty{4.50}{\milli\meter} $\times$ \qty{0.75}{\milli\meter} (\ac{PIC} and \ac{EIC} respectively).

\begin{figure}[h!]
\centering

\begin{subfigure}{\linewidth}
    \centering
    \includegraphics[width=\linewidth]{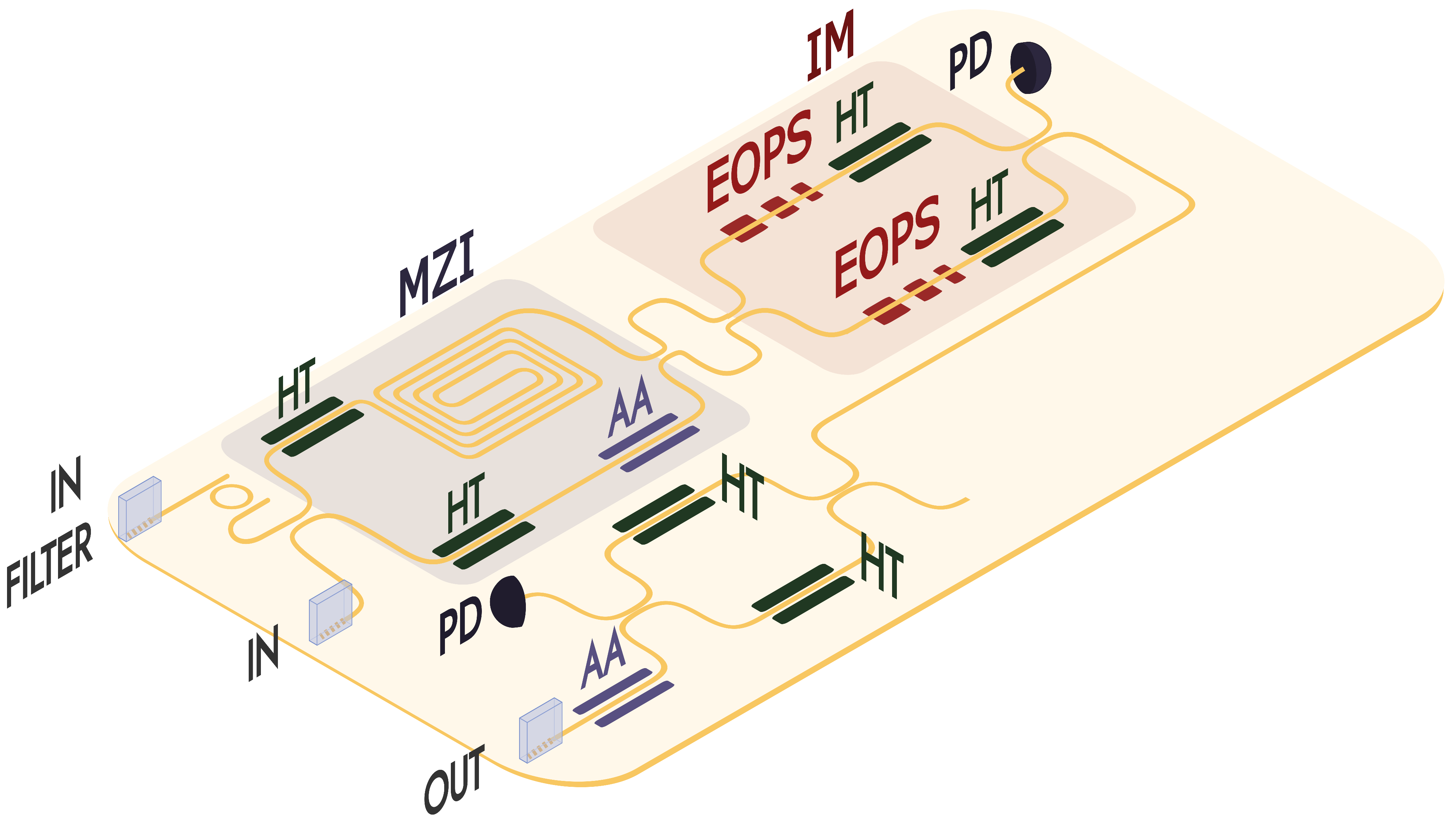}
    \caption{}
    \label{fig:2_3_PICalice}
\end{subfigure}

\begin{subfigure}{0.46\linewidth}
    \centering
    \includegraphics[width=\linewidth]{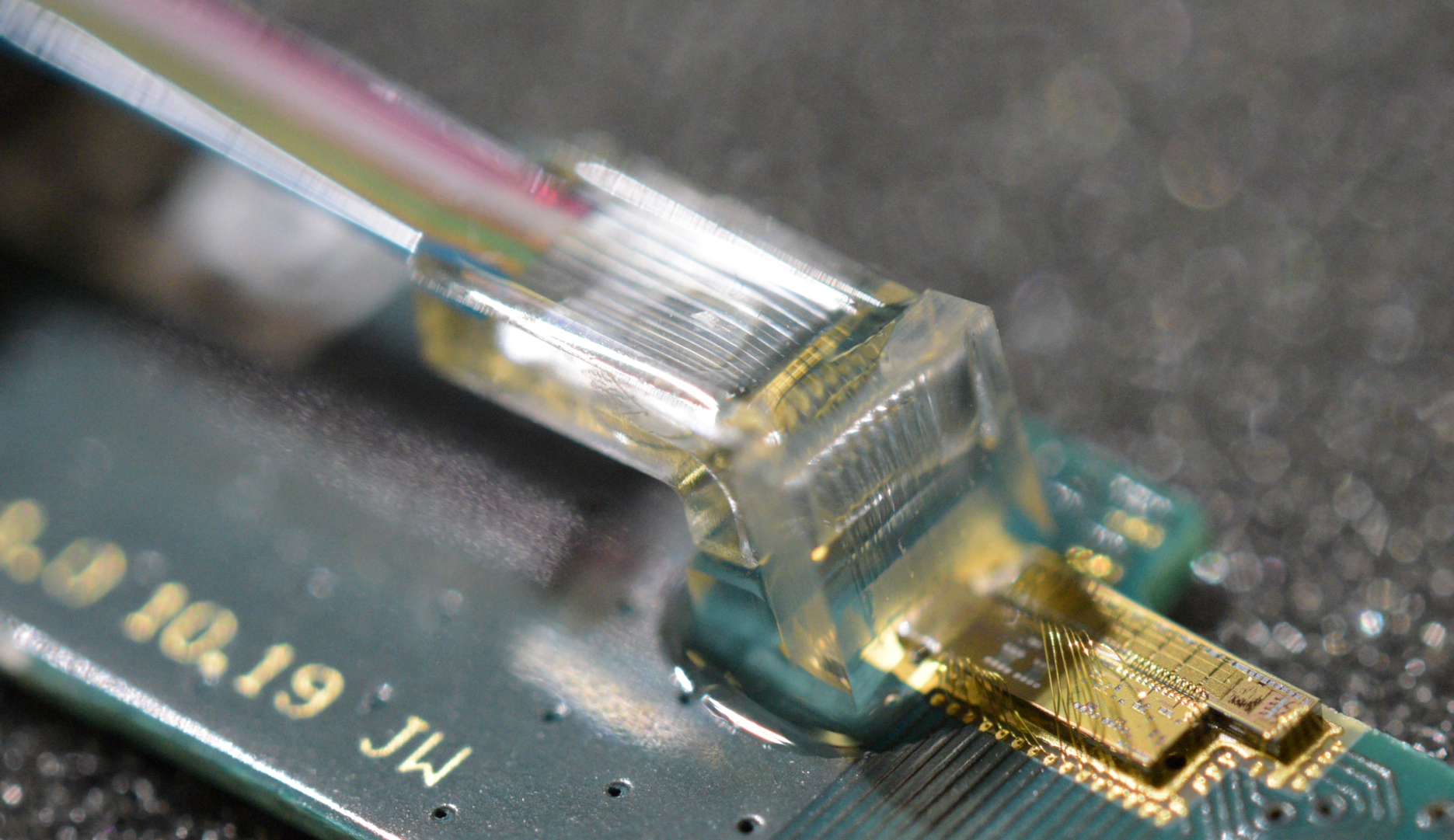}
    \caption{}
    \label{fig:2_3_2_GRATINGCOUPLER}
\end{subfigure}
\hspace{0.04\linewidth}
\begin{subfigure}{0.5\linewidth}
    \centering
    \includegraphics[width=\linewidth]{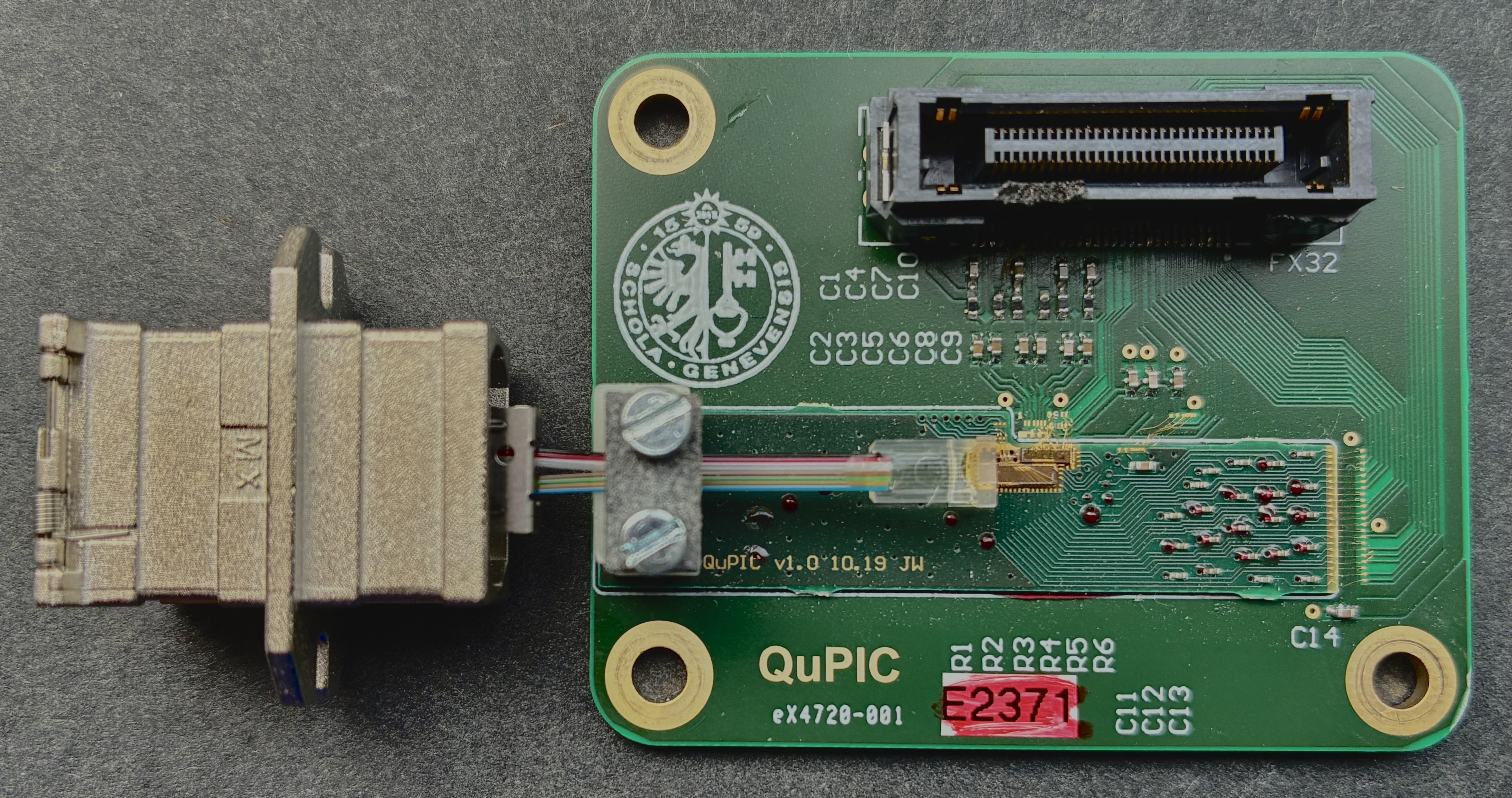}
    \caption{}
    \label{fig:2_3_2_PICEIC}
\end{subfigure}

\caption{a) Schematic of Alice's \ac{PIC}: HT – Heaters; EOPS – electro-optic phase shifters; MZI – Mach-Zehnder interferometer; IM – intensity modulator; AA – absorption attenuator; PD – photodetectors. b) Optical coupling with \ac{PIC}. c) \ac{PIC} interposer PCB and SPI connector.}
\end{figure}

The two chips are mounted on a custom interposer \ac{PCB}. The interposer is glued and bonded to a larger \ac{PCB} that interfaces with the remaining controlling electronics via a \textit{PCIe X4} connector Fig. \ref{fig:2_3_2_PICEIC}. The \ac{EIC} is directly bonded to the \ac{PIC} and integrates multiple SPI-controlled, high-speed differential driver channels that provide digitally tunable gain, bandwidth equalization, and up to \qty{3}{\volt}$_{pp}$ output swing.
The interposer also supports a 90-degree fiber array for optical coupling as seen in Fig. \ref{fig:2_3_2_GRATINGCOUPLER}. 
This light is coupled from this fiber array to and from chip using a grating coupler. An in-line polarization controller ($7.5 \times 2.5 \times 4.0$\,cm), upstream of the chip, aligns the incoming polarization to maximize coupling efficiency.

The transmitter \ac{PIC}, shown in Fig. \ref{fig:2_3_PICalice}, consists of four main functional blocks: a filtering stage comprised of a ring resonator on-chip, an unbalanced \ac{MZI} for qubit generation, an \ac{IM} for state encoding and a variable attenuator for mean photon number setting:
\begin{description}[leftmargin=0pt]

    \item[Spectral Cleaning] A ring resonator on chip, in a filtering configuration, is a wavelength-selective optical cavity coupled to two straight bus waveguides, an input bus and an output \textit{drop} port. Our chip's resonant wavelength and linewidth were measured to be $\lambda_{res}=1550.9$ and \qty{0.170}{\nano\meter} respectively. The frequency linewidth was characterized by sweeping an optical filter (\textit{JDSU TB9}) across the laser spectrum while measuring the \ac{SNSPD} count rate. 
    
    \item[Qubit Generation] The unbalanced \ac{MZI} generates the coherent \textit{early-late} pulse pairs with a \qty{400}{\pico\second} delay. This delay is achieved via a geometrical path-length difference of approximately $\Delta L\approx 3$ \unit{\centi\meter} between the two arms ($\Delta t= n_g \Delta L/c$, with a waveguide group index of $n_g\approx4.0$). The \ac{MZI} architecture consists of an input $2\times2$ \ac{BS} (one filtered input and one directly connected to the to the grating coupler), the unbalanced delay arms, and an output $2\times2$ \ac{BS}.
    \acp{TOPS} (\textbf{HT} in Fig. \ref{fig:2_3_PICalice}) in both arms actively control the relative phase required to maintain stable interference at the receiver, while an attenuator, in the short arm, compensates for propagation losses accumulated in the long arm, improving visibility. The \acp{TOPS} rely on the thermo-optic effect to modulate the refractive index of the waveguide, whereas the \acp{AA} exploit the free-carrier absorption effect to provide electrically controllable optical attenuation.  All the aforementioned \acp{BS} used on chip are Multimode Interference-based.

    \item[State encoding] The \ac{IM} encodes the quantum states at \qty{2.5}{\giga\hertz} (twice the laser's repetition rate). It uses a balanced \ac{MZI} with three \acp{EOPS} in each arm in a push-pull configuration. The \acp{EOPS} are driven jointly to produce the four required amplitude levels ($\mu_0$,  $\mu_0/2$, $\mu_0/4$, $\emptyset$). Additionally, \acp{TOPS} in each arm stabilize the \ac{IM} operating point.
    
    \item[Mean photon number setting] A final \ac{MZI} with \acp{TOPS} in both arms, attenuates the output to the desired mean photon number.

\end{description}

\begin{figure}[h]%
\centering
    \begin{subfigure}{1\linewidth}%
        \includegraphics[width=1\linewidth]{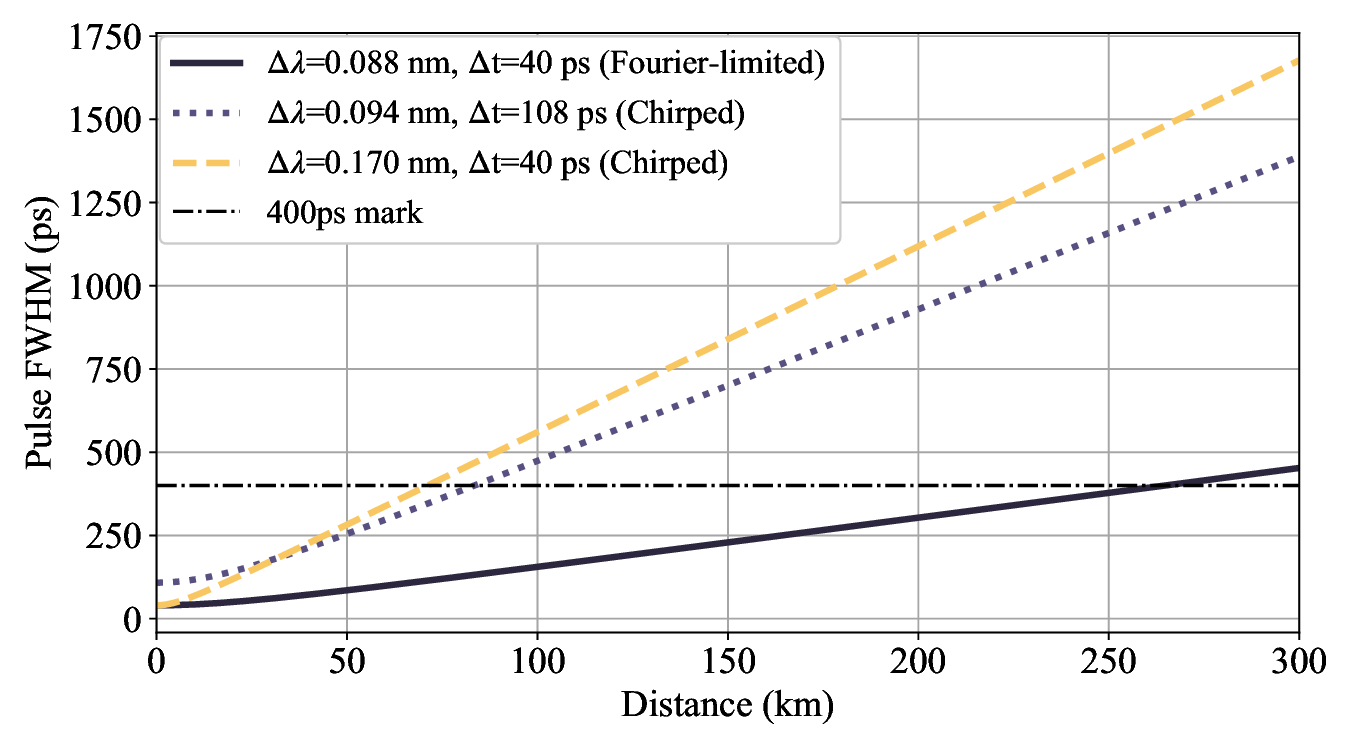}
        \caption{}
        \label{fig:chromatic_dispersion_simulation}
    \end{subfigure}
    \hfill

    \begin{subfigure}{1\linewidth}%
        \includegraphics[width=\linewidth]{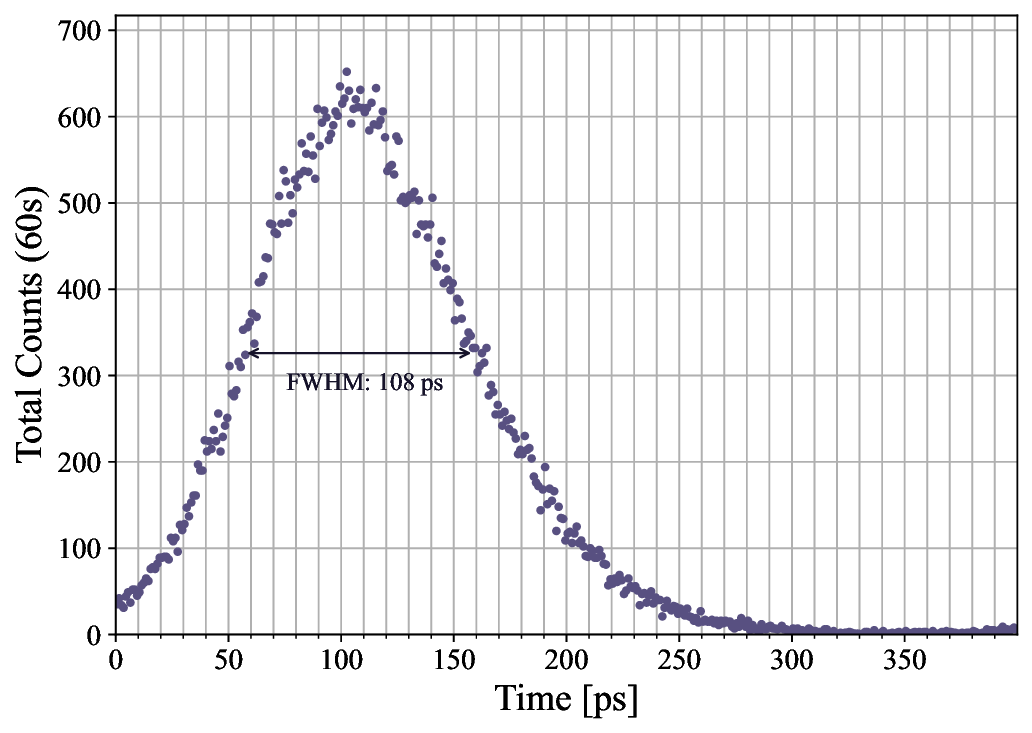}
        \caption{}
        \label{fig:laser_pulse}
    \end{subfigure}%
    
    \captionsetup{width=1\linewidth}

    \caption{a) Simulated pulse broadening due to chromatic dispersion effects, based on measured laser parameters [except the calculated \textit{Fourier-limited} scenario].     
    The 400 ps horizontal line indicates the time-bin width. The labels () indicate the function used for each curve. The laser's frequency linewidth was measured using the same method as the ring filter. b) Time-resolved intensity profile of the signal pulses measured with a \ac{SNSPD} (\qty{50}{\pico\second} jitter) and a \ac{TDC} (\textit{ID1000} (\qty{1}{\pico\second} resolution))
    }
 
\end{figure}

To assess the impact that halving the system's repetition rate had on chromatic dispersion sensitivity, we simulated the pulse broadening through a standard single-mode fiber with a dispersion parameter ($D$) of \qty{17.0}{\pico\second\per\nano\meter\per\kilo\meter}\cite{RAMASWAMI2010289}.
The solid line in Fig.\ref{fig:chromatic_dispersion_simulation} simulates the Fourier-limited scenario. In practice, the laser has an initial temporal \ac{FWHM} ($T_{FWHM_0}$) of \qty{40}{\pico\second} at the output of the chip, and a measured spectral width of $\Delta\lambda=$\qty{0.170}{\nano\meter} (after spectral filtering).
To model the impact of this chirp on the pulse propagation, we treat the signal as a chirped Gaussian pulse with an intensity half-width at $1/e$ of $T_0=T_{FWHM_0}/(2\sqrt{\ln 2})$. The temporal half-width $T_1(z)$ of the pulse after traveling a distance $z$ in the \ac{QC} is then given by:

\begin{equation}
\label{eq:temporal_broadening_output_pulse_width_chirped}
T_1(z) = T_0\left[\left(1+\left(\frac{C\cdot z}{L_D}\right)\right)^2+\left(\frac{z}{L_D}\right)^2\right]^{\frac{1}{2}},
\end{equation}
where $L_D=T_0^2/|\beta_2|,$ is the dispersion length (with $\beta_2=-\lambda^2D/(2\pi c)$ being the group velocity dispersion parameter), and $C$ is the chirp parameter derived from the measured spectral and temporal widths via $2\pi c/{\lambda_0}^2\cdot\Delta\lambda = (1+C^2)^{1/2}/T_0$.
At an \qty{0.170}{\nano\meter} spectral width, the broadened \ac{FWHM} would reach approximately \qty{560}{\pico\second} at \qty{100}{\kilo\meter} (Fig.\ref{fig:chromatic_dispersion_simulation}, dashed curve), significantly exceeding the \qty{400}{\pico\second} time-bin width, making working at these distances without the use of \ac{DCF} nonviable. By tuning the laser's bias voltage and RF driving signal, we optimized the operating point to narrow the spectrum to \qty{0.094}{\nano\meter} while increasing $T_{FWHM_0}$ to \qty{108}{\pico\second} (Fig.~\ref{fig:laser_pulse}). Under these optimized conditions, the broadened \ac{FWHM} remains approximately \qty{473}{\pico\second} (Fig.\ref{fig:chromatic_dispersion_simulation}, dotted curve).
While this \qty{473}{\pico\second} width still exceeds the \qty{400}{\pico\second} separation between early and late time-bins (dashed-dotted curve in Fig.\ref{fig:chromatic_dispersion_simulation}), key extraction remained possible at \qty{105}{\kilo\meter} due to the time-filter implemented in the receiver.

The detection events are time-stamped by the \ac{FPGA}, sampling at \qty{10}{\giga\hertz}, resulting in a fundamental temporal resolution of \qty{100}{\pico\second}, with one qubit mapped across eight of these \qty{100}{\pico\second} bins. Only detections falling into specific pre-determined bins corresponding to the \textit{early}-\textit{late} time-bins, are assigned to the raw key. The counts falling in the remaining \qty{100}{\pico\second} bins are discarded. This prevents "spill-overs" caused by detector jitter and pulse broadening from contributing to the \ac{QBER}.

The receiver consists of the integrated \ac{PIC} comprising the optical components (except detectors); two single photon detectors, one for each measurement basis; a control \ac{PCB} for thermal regulation, detector biasing, and signal readout; an \ac{FPGA} for real-time data acquisition and classical communication with the transmitter; and a computer for post-processing tasks including error correction and privacy amplification. The computer also interfaces with a microcontroller-based current supply for the receiver \ac{PIC}'s tunable beam splitter.


While \acp{SNSPD} achieve the best noise and time-resolution performance, their bulky and costly cryogenic systems makes them unsuitable for our deployable prototype. An attractive option to use in synchronous protocols are high-speed gated InGaAs/InP \acp{SPAD}. However, high-speed gating at gigahertz frequencies requires complex self-differencing or filtering architectures that are highly specific to an individual diode's characteristics\cite{Comandar2015-gx,10.1117/12.862118,He2023-fp}.
Free-running \acp{NFAD} therefore represent the best procurable option. For this work we used Princeton \acp{NFAD} peltier-cooled to -50$^\circ$C, which offer low jitter and dark count rates. The primary limitation, when using \acp{NFAD} is the dead time, which must be tuned to mitigate the effects of afterpulsing without hindering the performance of the protocol, since a long dead time significantly limits the maximum achievable raw key rate.


\begin{figure}[h!]
\centering
\includegraphics[width=0.95\linewidth]{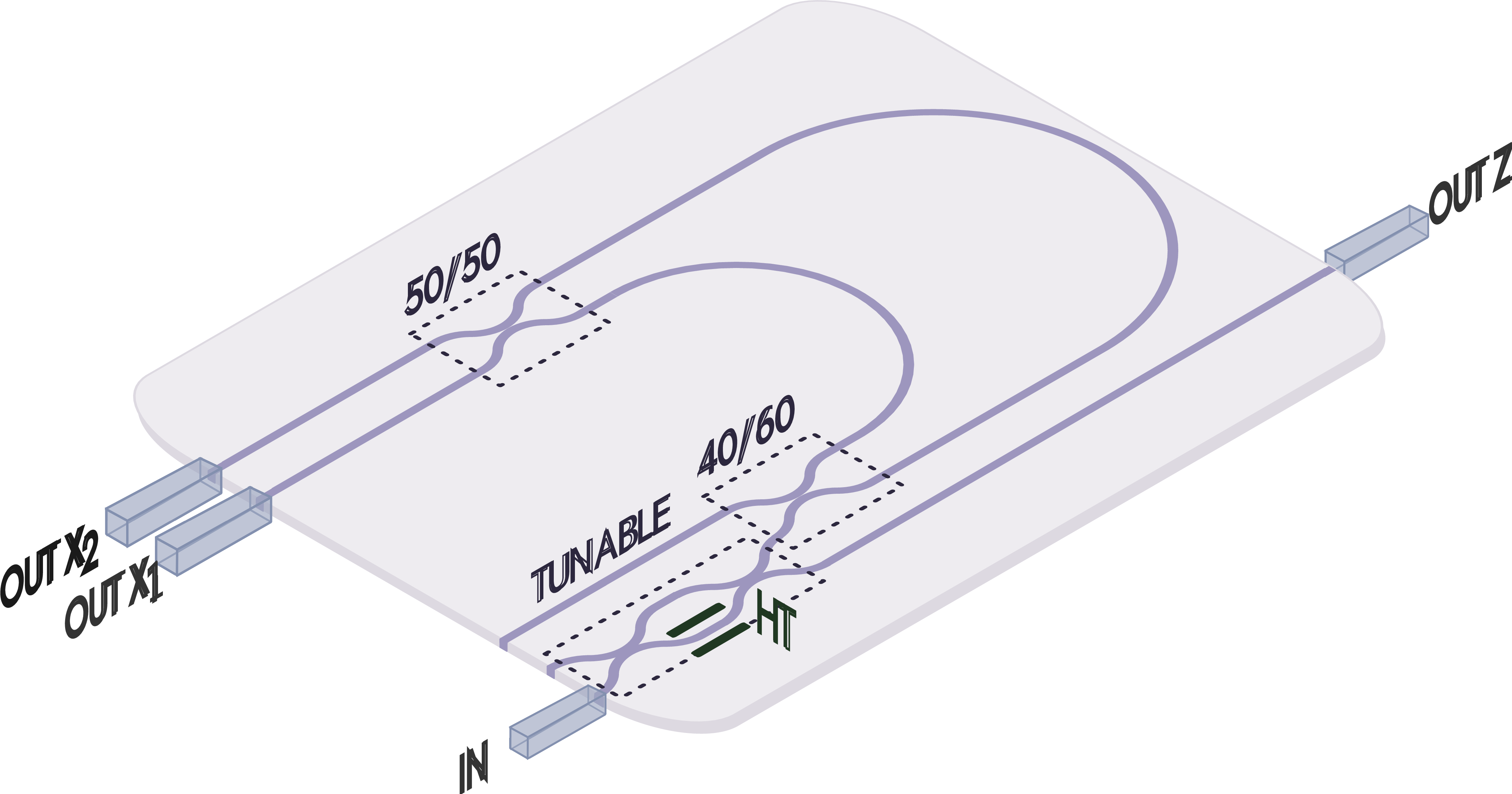}
\caption{Schematic of the receiver's \ac{PIC}: HT - heater, OUT X$_n$ - output connected to X basis detector, OUT Z - output connected to Z basis detector. 40/60 and 50/50 refer to the splitting ratio of the unbalanced interferometer input and output coupler respectively.}
\label{fig:BOB_PIC_Detail}
\end{figure}

It is crucial to have a low-loss receiver in order to achieve high key generation rates, therefore, with this loss budget in mind, the \ac{PIC} was fabricated through Femtosecond Laser Micromachining (FLM) in borosilicate glass (see Fig. \ref{fig:BOB_PIC_Detail}), where tightly focused ultrashort pulses directly write waveguides inside the substrate \cite{Corrielli2021FLM}. 

The FLM platform provides system-level robustness against \ac{SOP} variations through the intrinsic properties of its waveguides and couplers. Polarization-independent splitting ratios are ensured by the coupler geometry, while the inherent waveguide birefringence is mitigated during fabrication by inscribing stress tracks near the core\cite{Fernandes2012, Bhardwaj2004, McMillen2016}, exploiting the stress–optic effect. This ensures the interference visibility remains stable regardless of channel polarization fluctuations.

Although FLM offers less miniaturization than standard lithographic technologies, it provides high modularity, true 3D routing, and low-loss. The laser-written waveguides are well matched to optical fibers in both mode size and refractive index profile, enabling efficient fiber-to-chip coupling.
The total optical loss (taking coupling into account) in our \ac{PIC} is measured to be \qty{2.17}{\deci\bel} in the Z basis and \qty{3.52}{\deci\bel} in the X  basis.
Despite having an interferometer twice as long as the one reported by Sax\cite{Sax:23}, the low-loss nature of the FLM-written waveguides and high coupling efficiency, ensure only a marginal \qty{0.02}{\deci\bel} increase in loss within the X-basis arm.

Fig.\ref{fig:BOB_PIC_Detail} shows the schematic of the receiver's \ac{PIC} that can be divided into two main blocks one for basis selection and one for state measurement. 

\begin{description}[leftmargin=0pt]
    \item[Basis selection] A tunable \ac{BS} sets the measurement basis splitting ratio ($p_X^{(B)}$). This \ac{BS} is implemented as a symmetric \ac{MZI} with a microheater on one arm to enable thermo-optic phase tuning.  The microheater is fabricated by patterning the resistor via selective femtosecond laser ablation of a 100 nm Cr/Au film previously deposited on the substrate \cite{Ceccarelli2019TPS}. The splitting ratio can be continuously tuned from $0$ to $100\,\%$ in the X basis by applying \qty{63}{\milli\ampere} to the microheater.
    This tunable \ac{BS} provides two key advantages: it enables optimization of basis selection probabilities for different channel losses, making a single receiver design suitable for varied deployment scenarios; and it eliminates the need for multiple specialized chip designs during fabrication, reducing production complexity and cost while improving scalability.
 
    \item[State measurement] The $Z$-basis arm of the tunable \ac{BS} is connected directly to the output of the \ac{PIC}, while the $X$-basis arm connects to an unbalanced receiver \ac{MZI} for the time-bin interferometric measurement. The receiver \ac{MZI} features a path-length delay matched to Alice's transmitter within \qty{0.01}{\pico\second}, measured via whitelight interferometry using an \textit{EXFO IQ-203 Optical Test System} equipped with a PMD module.
    Because the long waveguide arm introduces an additional expected propagation loss of approximately \qty{1.7}{\deci\bel}, the receiver unbalanced \ac{MZI} is passively designed with a fixed 40/60 input directional coupler to bias more power into the long arm and offset its attenuation. However, due to fabrication tolerances, a residual power imbalance of $2.5\,\%$ remains.
    A minimum fringe visibility of $99.73\,\%$ (evaluated under the worst-case incoming polarization state) was measured with the \ac{PIC} temperature-stabilized at \qty{14.2}{\degreeCelsius}. The visibility's baseline contribution to the phase error rate ($\phi_z$) is given by $(1-V)/2$, making its impact negligible under these operating conditions.
\end{description}



To assess the viability of our system for real-world deployment, we characterized its performance in the Geneva metropolitan network. The transmitter and receiver were located in two different University buildings connected by 4.6 km of installed fiber (\qty{9.4}{\deci\bel} total loss). Initially, the detectors were operated at \qty{-50}{\degreeCelsius} with standard Peltier cooling.
An automatic feedback loop stabilized the relative phase between the transmitter and receiver by tuning the phase in the transmitter's \ac{MZI} based on the measured value of the phase error rate ($\phi_z$). This was implemented even though the stability of the paired \acp{MZI} exceeded 40 minutes without active stabilization under deployment conditions, in anticipation of long term operation.

\begin{figure}[h]
    \centering
    \includegraphics[width=0.5\textwidth]{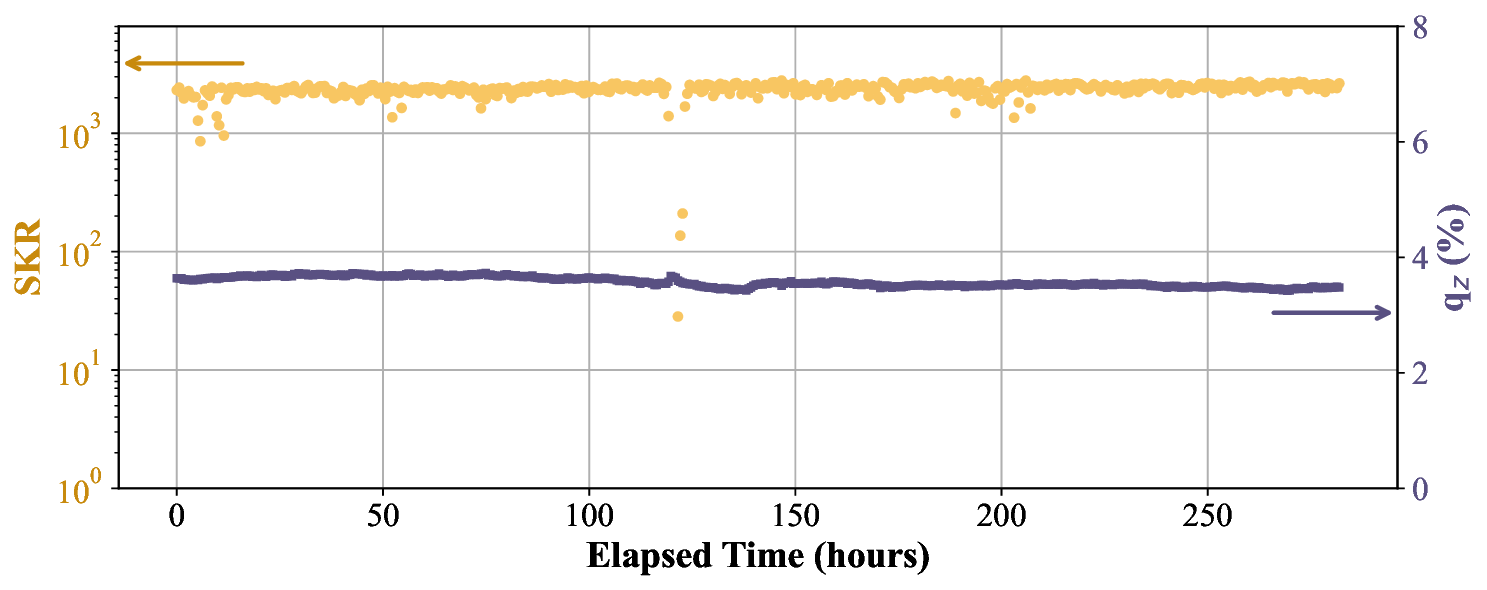}
    \caption{Stability of the \ac{SKR} and q$_z$ over a 282h (around 12 days) measurement of a point-to-point key exchange in the deployed fiber link.}
    \label{fig:2_4_Pinchat_point_to_point_stability}
\end{figure}

The experimental results and settings are summarized in Table \ref{tab:experimental_parameters}, Table \ref{tab:SKR_FIBER} and Fig. \ref{fig:2_4_Pinchat_point_to_point_stability}. For error correction, we perform in real-time a Cascade algorithm with a block size of 8192 bits and 1.05 efficiency.

\begin{table}[h!]
 \captionsetup{width=1\linewidth}
  \centering
\setlength{\tabcolsep}{5pt}
\renewcommand{\arraystretch}{2}
\begin{tabular}{|c||c|c|c|c|}
\cline{2-5} 
\multicolumn{1}{c|}{} & $\mu_0$ & $\mu_1$ & $p_x$ & $p_z$ \\
\cline{2-5}
\noalign{\vskip 2pt}
\hline
\makecell{Short \\ Distances}&0.50 & 0.26 & 0.2 & 0.80\\
\hline
\makecell{Long \\ Distances}&0.50 & 0.26 & 0.34 & 0.66\\
\hline
\end{tabular}
\caption{Experimental parameters. $\mu_0$ and $\mu_1$ are the mean photon number of the signal and decoy states respectively and the ratios of sending $x$ and $z$ states are $p_x$ and $p_z$ respectively. Short Distances: $\leq30$\unit{\kilo\meter}; Long Distances: $\geq30$\unit{\kilo\meter} $\&$ $\geq35$ \unit{\deci\bel} .}
 \label{tab:experimental_parameters}
\end{table}

As shown in the first entry of Table \ref{tab:SKR_FIBER}, we measured a \ac{SKR} of 6.7 kbps with a $\phi_z$ of only $2.0\,\%$. The robustness of the system was further validated by continuous operation over 12 day-and-night cycles, over the \qty{4.6}{\kilo\meter} installed metropolitan link, without human intervention (Fig. \ref{fig:2_4_Pinchat_point_to_point_stability}). The system was still stable and fully operational when the experiment was terminated, indicating that the duration of the key exchange was limited only by our experimental schedule rather than by any inherent instability in the hardware or the phase compensation feedback loop. The \ac{SKR} drop seen around hour 120 in Fig. \ref{fig:2_4_Pinchat_point_to_point_stability} is attributed to a temporary loss of phase locking in the stabilization loop, from which the system autonomously recovered without manual intervention.
Following the stability demonstration, we characterized the system's performance limits by extending the link with additional standard \ac{SMF} spools at the receiving site.
Initially, using Peltier cooling at \qty{-50}{\degreeCelsius}, the detectors exhibited a \ac{DCR} of more than $1.4\times10^3$ cps. At the \qty{59.8}{\kilo\meter} mark (\qty{-17.0}{\deci\bel}), this noise level resulted in \acp{QBER} close to $10\,\%$ and a modest \ac{SKR} of $0.8$ kbps. Under these conditions, the dark counts contribute significantly to the total detection events, limiting the maximum secure distance.
By transitioning to Stirling cooling at \qty{-85}{\degreeCelsius}, the noise floor was reduced by an order of magnitude, consequently,  for the same \qty{59.8}{\kilo\meter} link, the $q_z$ dropped to $2.9\,\%$ and the \ac{SKR} increased to $5.6$ kbps, a direct result of lowering the detector noise floor.

\newcolumntype{"}{@{\hskip\tabcolsep\vrule width 1pt\hskip\tabcolsep}}[h]
\makeatother
\setlength{\tabcolsep}{5pt}
\begin{table*}[t]
\captionsetup{width=.98\textwidth,justification=justified,singlelinecheck=false, format=plain} 
\centering
\noindent 
\begin{tabular}{|c|c"c|c|c|c|c|c|c|c"c|c"c|c|c|}
\hline
\hline
\multicolumn{2}{|c"}{Quantum Channel} & \multicolumn{6}{c"}{Detectors} & \multicolumn{2}{c"}{Distillation} & \multicolumn{3}{c|}{Performance}
\\ \hline
\hline
\makecell{L \\ (\unit{\kilo\meter})}
& \makecell{Att \\ (dB)} & $ \makecell{\eta_{data}\\(\%)}$ & $\makecell{\tau_{data}\\(\unit{\micro\second})}$ & $\makecell{DCR_{data}\\(cps)}$ & $\makecell{\eta_{mon}\\(\%)}$ & $\makecell{\tau_{mon}\\ (\unit{\micro\second})}$ & $\makecell{DCR_{mon}\\(cps)}$  & \makecell{T\\ ($^\circ$ C)} & $p_X^{(B)}$ & $n$ & $\makecell{t_{ccp} \\(s)} $& $\makecell{\mathrm{q}_z \\ (\%)} $ & $\makecell{\phi_z \\ (\%)} $& \makecell{ SKR \\ (kbps)}\\

\hline
\hline
4.6 & 9.4 &  20 & 24 & $2.7\times10^3$ & 20 & 28 & $1.4\times10^3$ & -50 & 0.50 & $8\times10^6$& 7.2$\times10^3$ & 3.3 & 2.0 & 6.7 \\

\hline

59.8  & 17.0 & 20 & 24 & $2.7\times10^3$ & 20 & 28 & $1.4\times10^3$ & -50 & 0.50 & $8\times10^6$ & $1.7\times 10^5$ & 9.7 & 9.9 & 0.8 \\

\hline
\hline

17.3 & 12.0 & 20 & 4 & \num{1.0e3} & 19 & 12  & \num{2.4e2} & -85 &0.10 & $8\times10^6$ & $1.6\times10^3$ & 1.9 & 3.6 & 13.2\\
\hline
29.9 & 14.9 & 21 & 12 & \num{3.7e2} & 20 & 28 &  \num{1.3e2} & -85 & 0.10 & $8\times10^6$ & $2.0\times10^3$ & 2.5 & 3.8 & 9.8 \\
\hline
59.8 & 17.0 & 21 & 12 & \num{2.0e2} &  20 & 28 & \num{1.3e2} & -85 & 0.10 & $8\times10^6$ & 3.3$\times10^3$ & 2.9 & 5.1 & 5.6 \\
\hline
105.4 & 33.4 & 21 & 24  & \num{1.0e2}& 20 & 32 & \num{1.2e2} & -85 & 0.35  &  $8\times10^4$ & $1.8\times10^3$ & 7.4 & 10 & 0.3 \\

\hline
\hline
4.6 & 35 & 21 & 24 & \num{1.0e2} & 20 & 28 & \num{1.3e2} & -85 & 0.1 & $8\times10^6$ & $1.7\times10^4$ & 1.3 & 6.0 & 0.9  \\
\hline
4.6 & 40 & 19 & 8 &  \num{7.4e1} & 20 & 105 & \num{7.7e1} & -85 & 0.4 & $8\times10^4$ & $6.7\times10^4$ & 3.3 & 13.6 & 0.16 \\

\hline

\end{tabular}

\caption{Experimental parameters and performance metrics for the metropolitan link with varying total fiber lengths. All measurements were performed on an installed metropolitan network, with additional standard \ac{SMF} spools used to evaluate system performance at extended distances (lines 2-6), or extra attenuation (lines 7-8). The data highlights the transition from Peltier cooling at \qty{-50}{\degreeCelsius} to Stirling cooling at \qty{-85}{\degreeCelsius} to overcome \ac{DCR} limitations at higher attenuation. $L$ refers to the \ac{QC} length, $Att$ to the total attenuation in the \ac{QC}, $\eta_x$ and $\tau_x$ refer to the efficiency and dead time, respectively, of the $data$ and $monitor$ detectors used. $T$ is the detector's operating temperature, and $p_X^{(B)}$ denotes Bob's probability of measuring in the $X$ basis. $n$ and $t_{ccp}$ are the raw key block size and acquisition time respectively, $q_z$, $\phi_z$ and $SKR$ denote the $\ac{QBER}$ in the Z basis, phase error rate and Secret Key Rate.} 
\label{tab:SKR_FIBER}

\end{table*}

To isolate the effects of noise arising from chromatic dispersion, we performed measurements using a \ac{VOA} (\textit{EXFO FVA-3100 series}) at the end of the \qty{4.6}{\kilo\meter} metropolitan link, to set the \ac{QC} to the desired loss. At \qty{-35}{\deci\bel}, the system maintains a low $q_z$ ($1.3\,\%$), however, at \qty{-40}{\deci\bel}, the signal count rate becomes comparable to the \ac{DCR}. This, aligned with the bias on $p_X^{B}=0.4$, causes $\phi_z$ to spike from $6\,\%$ to $13.6\,\%$. This establishes \qty{40}{\deci\bel} as the fundamental loss budget of our system; beyond this loss, the noise prevents the extraction of a positive key rate.

Comparing these \ac{VOA} results with the long-distance fiber measurements reveals the significant impact of chromatic dispersion. Despite experiencing \qty{6.6}{\deci\bel} more loss, the \ac{VOA} configuration exhibits a $q_z$ more than $2\times$ lower, demonstrating that chromatic dispersion becomes the dominant performance limitation at longer distances.

Looking into the trade-off between raw repetition rate  (\qty{2.5}{\giga\hertz} in\cite{Sax:23} and \qty{1.25}{\giga\hertz} in this work) and key quality, we observed consistently lower \ac{QBER} values compared to those reported by\cite{Sax:23} for similar attenuation levels. 
Comparing the performance at \qty{-30}{\deci\bel}, our system achieved a $q_z$ of $1.3\,\%$ and an SKR of 1.80 kbps, whereas the system in \cite{Sax:23}, despite doubling our repetition rate, demonstrated a SKR of 2.9 kbps with a significantly higher \ac{QBER} of $3.6\,\%$. This trend of high-quality key extraction continues through higher attenuation levels.
While several system improvements beyond the clock rate contribute to this performance, the wider qubit time-bins inherently reduce the sensitivity to chromatic dispersion and detector jitter. This suggest that for similar systems, optimizing for a low \ac{QBER} floor with a moderate repetition rate can be more effective for secret key extraction than simply increasing the clock rate, particularly for short distance links where detector saturation is expected.

In this work, an integrated photonics based \ac{QKD} prototype has been successfully evaluated in a metropolitan network.
By integrating all components into optical chips (except the laser and detectors) and consolidating the totality of the \ac{QKD} transmitter and receiver into standard 19-inch racks, we demonstrated a stable, real-world architecture capable of autonomous key exchange. The system demonstrated over 280 hours of stable, autonomous key exchange over an installed \qty{4.6}{\kilo\meter} fiber link. Complementary characterization using extra fiber spools and attenuation demonstrated the system's resilience to loss and dispersion, where we observed positive key rates up to \qty{105.4}{\kilo\meter}, and separately, \qty{40}{\decibel} of total attenuation.
This demonstration represents a significant advancement toward the practical viability of chip-based \ac{QKD} systems. Our results suggest that \ac{PIC}-based architectures provide a promising pathway to streamlining the implementation of quantum security within standard telecommunications networks.

\begin{acknowledgments}

We gratefully acknowledge Sicoya GmbH for the helpful discussion regarding the photonic integrated chip used in the transmitter's setup. We thank ID Quantique for providing the chassis, PC/FPGA system for Alice. We thank Claudio Barreiro and for his contributions to the \ac{PCB} design work and David Cabrerizo for helping with the \ac{PCB} assembly. We would also like to acknowledge Shashank Kumar for his contributions to the fiber network setup.
\end{acknowledgments}

\section*{Funding}
This work was supported by the Swiss State Secretariat for Research and Innovation (SERI) (Contract No. UeM019-3) and European Quantum Flagship project openQKD (857156); R.O would like to acknowledge funding by PNRR MUR project PE0000023-NQSTI (Spoke 7)

\section*{Data Availability Statement}
The data that support the findings of this study are available from the corresponding author upon reasonable request.

\bibliography{aipsamp}

\end{document}